\documentclass[a4paper]{PoS}

\title{Baryon spectroscopy in the unquenched quark model}

\ShortTitle{Baryon spectroscopy in the unquenched quark model}

\author{\speaker{Roelof Bijker}, Gustavo Guerrero-Navarro and Emmanuel Ortiz-Pacheco\\
Instituto de Ciencias Nucleares, UNAM, AP 70-543, Mexico DF, Mexico\\
E-mail: \email{bijker@nucleares.unam.mx},\\
\email{gustavohazel@ciencias.unam.mx},\\
\email{emmanuelo@ciencias.unam.mx}}

\abstract{We discuss some applications of the unquenched quark model which is an 
extension of the CQM that includes the effects of sea quarks via a $^{3}P_{0}$ quark-antiquark 
pair-creation mechanism. Particular attention is paid to the electromagnetic couplings and $\beta$ 
decays of baryons. It is shown that the observed discrepancies between the experimental data and 
the predictions of the CQM can be accounted for in large part by the effects of sea quarks in the 
unquenched quark model.}

\FullConference{XVII International Conference on Hadron Spectroscopy and Structure\\
                 25-29 September, 2017\\
                 University of Salamanca, Salamanca, Spain}

\newcommand{\ba}{\begin{eqnarray}}
\newcommand{\ea}{\end{eqnarray}}
\begin{document}

\section{Introduction}

The constituent quark model (CQM) describes the nucleon as a system of three constituent, 
or valence, quarks. Despite the successes of the CQM (e.g. masses, electromagnetic couplings, 
and magnetic moments), there is compelling evidence for the presence of sea quarks from other  
observables such as the observed flavor asymmetry of the proton, the proton spin crisis, 
and the systematics of strong decays of baryons.  

In the CQM, baryons are described in terms of a configuration of 
three constituent (or valence) quarks neglecting the effects of pair-creation (or 
continuum couplings). Above threshold these couplings lead to strong decays and below 
threshold to virtual higher-Fock components ($qqq-q\bar{q}$) in the baryon wave function. 
The effects of these multiquark configurations (or sea quarks) can be studied by unquenching 
the CQM. 

In this contribution, we study the importance of sea quarks for the electromagnetic 
and weak couplings of baryons.

\section{Unquenched quark model}

In the unquenched quark model (UQM), the effects of sea quarks are included via a $^{3}P_{0}$ 
quark-antiquark pair-creation mechanism \cite{tornqvist,zenc,baryons,uqm1,uqm2}. The pair-creation 
mechanism is inserted at the quark level and the one-loop diagrams are calculated by summing 
over a complete set of intermediate baryon-meson states. As a result, 
the baryon wave function is given by the sum of a contribution of the valence 
quarks $\mid A \rangle$ and a higher-Fock component consisting of an intermediate 
baryon-meson configuration $\mid BC \rangle$
\ba 
\mid \psi_A \rangle = {\cal N}_A \left\{ \mid A \rangle + \gamma 
\sum_{BC l J} \int d \vec{K} k^2 dk \, \mid BC,l,J; \vec{K},k \rangle \, 
\frac{ \langle BC,l,J; \vec{K},k \mid T^{\dagger}( ^{3}P_{0}) \mid A \rangle } 
{\Delta E_{A \rightarrow BC}(k)} \right\} ~,
\ea
where the energy denominator is calculated in the rest frame of baryon A 
\ba 
\Delta E_{A \rightarrow BC}(k) &=& m_A - \sqrt{m_B^2 + k^2} - \sqrt{m_C^2 + k^2} ~.
\ea
Here $\vec{k}$ and $l$ denote the relative radial momentum and relative orbital angular momentum of 
the baryon-meson system BC. The operator $T^{\dagger}$ creates a quark-antiquark pair in the 
$^{3}P_0$ state with the quantum numbers of the vacuum \cite{uqm1,uqm2,roberts} 

In the UQM, the matrix elements of an observable $\hat{\cal O}$ are calculated as 
$\langle \psi_A \mid \hat{\cal O} \mid \psi_A \rangle$ 
which is the sum of a contribution from the valence part and from the continuum component 
(or sea quarks). Previous studies have shown that whereas the effects of quark-antiquark pairs 
do not change the good results of the CQM for the magnetic moments \cite{uqm1}, they provide 
important contributions to observables like the orbital angular momentum in the spin of the 
proton \cite{uqm1}, the flavor asymmetry of the proton \cite{uqm2}, strangeness content of 
nucleon electromagnetic form factors \cite{baryons,uqm3}, strangeness suppression \cite{uqm4}, 
and self-energy corrections to baryon masses \cite{uqm5}. In the next section, we discuss 
the importance of higher-Fock components in electromagnetic and weak decays of baryons. 

\section{Results}

In this section, we discuss some recent results for electromagnetic and weak couplings. 
A more detailed account will be given in future publications \cite{gustavo,emmanuel}. 
In the present calculation, the sum over intermediate states is limited to octet and decuplet 
baryons in combination with pseudoscalar mesons. The contributions of radially excited baryons 
and mesons are not taken into account.  

\begin{table}
\centering
\begin{tabular}{cccc}
\noalign{\smallskip}
\hline
\noalign{\smallskip}
\hline
\noalign{\smallskip}
& CQM & UQM & Exp \cite{PDG} \\
\noalign{\smallskip}
\hline
\noalign{\smallskip}
$\Gamma(\Delta \rightarrow N \gamma)$ & 399 & 608 & $703 \pm 61$ \\
$\pi$ && 554 & \\
$\pi$, K && 582 & \\
$\pi$, K, $\eta$, $\eta'$ && 608 & \\
\noalign{\smallskip}
\hline
\noalign{\smallskip}
\hline
\end{tabular}
\caption{Radiative decay widths of the $\Delta$ resonance in keV.}
\label{emtab}
\end{table}

\subsection{Electromagnetic decays}

\begin{figure}[b]
\centering
\includegraphics[width=0.5\textwidth]{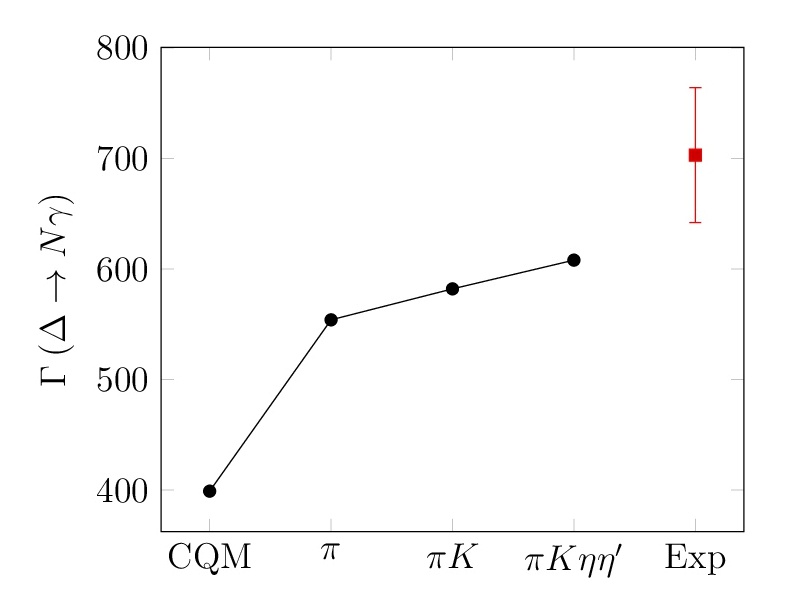}
\caption{Radiative decay widths of baryons in keV: $\Gamma(\Delta \rightarrow N \gamma)$.}
\label{emfig}
\end{figure}

Here we consider the effect of sea quarks for the electromagnetic decay of the $\Delta$ 
resonance. The radiative width for this process can be obtained in terms of a sum over 
helicity amplitudes 
\ba
\Gamma(\Delta \rightarrow N \gamma) = \frac{E_N p_{\gamma}^2}{2\pi m_{\Delta}} \sum_{\nu>0} | A_{\nu} |^2 ~,
\ea
which in turn for ground state baryons is related to the transition 
magnetic moment $\mu_{\Delta N}$ 
\ba
\sum_{\nu>0} | A_{\nu} |^2 = 4 \pi p_{\gamma} \, \mu_{\Delta N}^2 ~.
\ea
The results in Table~\ref{emtab} and Fig.~\ref{emfig} show that the coupling to the pion cloud 
accounts in great part for the observed discrepancy between the quark model value $399$ keV and 
the experimental value $703 \pm 61$ keV. The contribution of intermediate states containing 
kaons and $\eta$ mesons is small compared to that of the pions. 
 
\subsection{Weak decays}

Semileptonic decay processes of baryons $A \rightarrow A' + e^- + \bar{\nu}_e$ are 
described by the axial couplings which for octet baryons can be expressed in terms of 
$F$ and $D$. In the quark model, they are given by $F=2/3$ and $D=1$, and one obtains 
\ba
g_A(n \rightarrow p) &=& F+D = \frac{5}{3} ~, 
\nonumber\\
g_A(\Sigma^- \rightarrow n) &=& F-D = -\frac{1}{3} ~, 
\ea 
compared to the experimental values, $1.2701 \pm 0.0025$ and $-0.340 \pm 0.017$, respectively. 
In the Cabibbo approach, $F$ and $D$ are determined from the experimental axial couplings, 
$g_A(n \rightarrow p)$ and $g_A(\Sigma^- \rightarrow n)$, leading to effective values 
$F=0.465$ and $D=0.805$. With these values the semileptonic decay processes of baryons 
are described very well \cite{Cabibbo}. 

\begin{figure}
\includegraphics[width=0.5\textwidth]{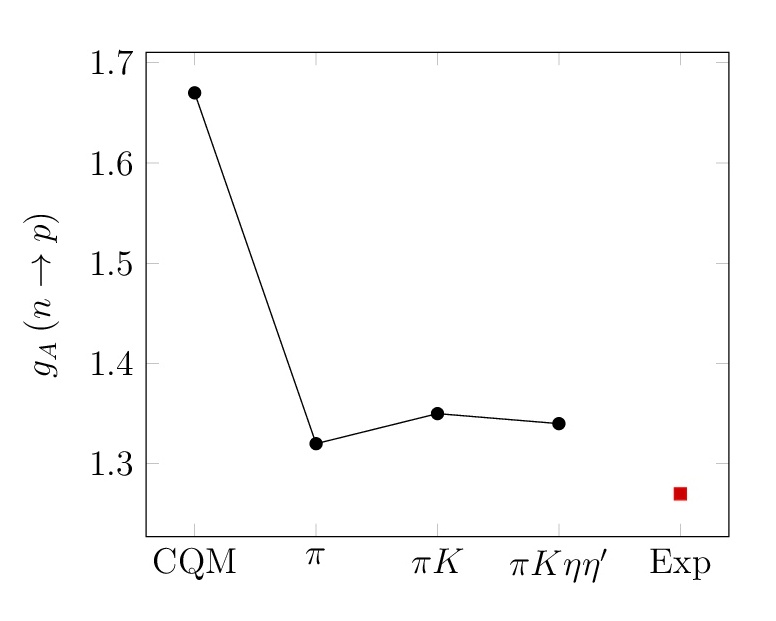}
\includegraphics[width=0.5\textwidth]{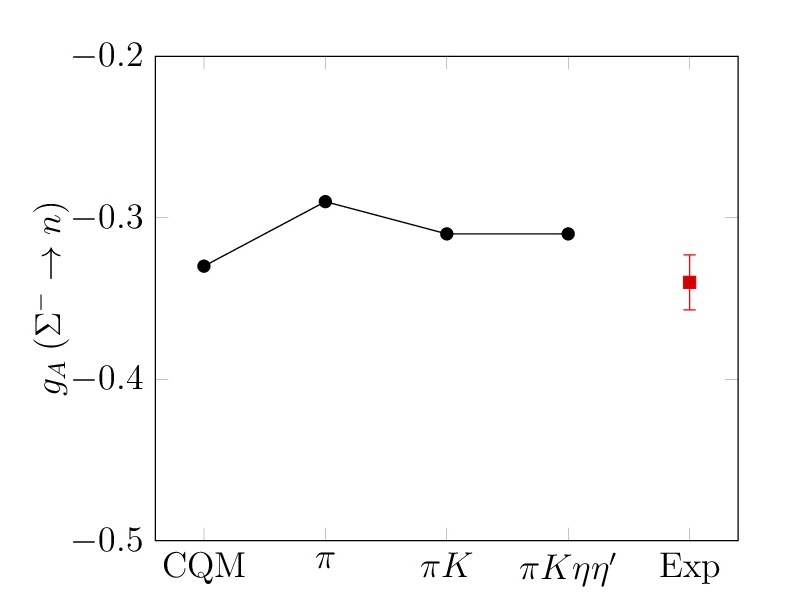}
\caption{Axial couplings of octet baryons: $g_A(n \rightarrow p)$ (left) 
and $g_A(\Sigma^- \rightarrow n)$ (right).}
\label{weakfig}
\end{figure}

In Table~\ref{weaktab} and Fig.~\ref{weakfig} we show a comparison of the results for the 
axial couplings in the CQM and the Cabibbo approach with those of the unquenched quark model. 
The contribution of the pions is responsable for a substantial lowering of the neutron axial 
coupling from the CQM value of $1.67$ to $1.32$ thus bringing it in much closer agreement 
with experiment without the need to introduce effective values of $F$ and $D$. 
On the other hand, the result for the $\Sigma^-$ hyperon which is described very well 
in the CQM is hardly changed. This shows that the effective values of $F$ and $D$ used in the 
Cabibbo approach can be accounted for in large part by the coupling to the pions. 
A similar conclusion was reached in an earlier study of the meson-cloud model \cite{Speth}. 

\begin{table}
\centering
\begin{tabular}{cccccccr}
\noalign{\smallskip}
\hline
\noalign{\smallskip}
\hline
\noalign{\smallskip} 
$g_A$ & \multicolumn{2}{c}{CQM} & UQM & Cabibbo & Exp \cite{PDG} \\
\noalign{\smallskip}
\hline
\noalign{\smallskip}
$n \rightarrow p$ & $F+D$ & $1.67$ & $1.34$ & $1.27$ & $1.2701 \pm 0.0025$ \\
\noalign{\smallskip}
$\pi$ &&& $1.32$ && \\
$\pi$, K &&& $1.35$ & \\
$\pi$, K, $\eta$, $\eta'$ &&& $1.34$ & \\
\noalign{\smallskip}
\hline
\noalign{\smallskip} 
$\Sigma^- \rightarrow n$ & $F-D$ & $-0.33$ & $-0.31$ & $-0.34$ & $-0.340 \pm 0.017$ \\
\noalign{\smallskip}
$\pi$ &&& $-0.29$ && \\
$\pi$, K &&& $-0.31$ & \\
$\pi$, K, $\eta$, $\eta'$ &&& $-0.31$ & \\
\noalign{\smallskip}
\hline
\noalign{\smallskip}
& $F$ & $\frac{2}{3}$ & $\frac{2}{3}$ & $0.465$ & \\
& $D$ & $1$           & $1$           & $0.805$ & \\
\noalign{\smallskip}
\hline
\noalign{\smallskip}
\hline
\end{tabular}
\caption{Axial couplings of octet baryons.}
\label{weaktab}
\end{table}

\section{Summary and conclusions}

In this contribution, we studied the importance of higher-Fock components (or sea quarks) 
in electromagnetic and weak decays of baryons in the framework of the unquenched quark model. 
It was shown that the observed discrepancies between the experimental data and the predictions 
of the CQM can be accounted for in large part by the contributions of quark-antiquark pairs 
in the UQM. In particular, for the decays $\Delta \rightarrow N + \gamma$ and 
$n \rightarrow p + e^- + \bar{\nu}_e$ the contribution of the sea quarks is dominated by the pions.   

\section*{Acknowledgements}
This work was supported in part by grant IN109017 from DGAPA-UNAM, Mexico
and grant 251817 from CONACyT, Mexico.

\end{document}